\documentclass{iopart}
\usepackage{amssymb}
\usepackage{iopams}
\usepackage[usenames]{color}
\usepackage{graphicx}
\usepackage{dcolumn}
\usepackage{bm}
\usepackage[sort&compress, numbers]{natbib}

\newcommand{\pc}{\pb_{c}}	
\newcommand{\pbc}{\pb_{c}^{b}}	
\newcommand{\pwc}{p_{c}^{w}}	
\newcommand{\pbinf}{P_\infty}		
\newcommand{\pwinf}{\bar{P}_\infty}		
\newcommand{\pz}{p_{0}}	
\newcommand{\pbz}{p_{0}^{b}}	
\newcommand{\pwz}{p_{0}^{w}}	
\newcommand{\pb}{p}	
\newcommand{\pw}{{\bar{p}}}	
\newcommand{\pestzeta}{\hat{p}}		
\newcommand{\pest}{p^*}		
\newcommand{\mec}{\chi}		
\newcommand{\mecw}{\bar{\chi}}		
\newcommand{\mecb}{\chi}		
\newcommand{\EC}{\mathcal{X}}		
\newcommand{\patternb}{P}		
\newcommand{\clusternob}{n}		
\newcommand{\clusternow}{\bar{n}}		
\newcommand{\perpolb}{g}		
\newcommand{\perpolw}{\bar{g}}		
\newcommand{\prefacb}{{a}}		
\newcommand{\prefacw}{{\bar{a}}}		
\newcommand{\lattice}{\Lambda}		
\newcommand{\matchlattice}{\bar{\Lambda}}		
\newcommand{\pdec}{{p_{dec}}}	
\newcommand{\dpq}{\Delta}	
\newcommand{\EQ}[1]{Equation~\ref{eq:#1}}

\newcommand{\FIG}[1]{Figure~\ref{fig:#1}}
\newcommand{\TAB}[1]{Table~\ref{tab:#1}}
\newcommand{\REF}[1]{ref.~\citep{#1}}
\def\newblock{\hskip .11em plus .33em minus .07em}

\begin{document}

\bibliographystyle{unsrtnat}

\title{Topological estimation of percolation thresholds}

\author{%
Richard A. Neher${}^1$\footnote{Present address: KITP, University of California, Santa Barbara.},
Klaus Mecke${}^2$
and 
Herbert Wagner${}^1$%
}

\address{$^1$ Arnold-Sommerfeld-Center for Theoretical Physics, LMU M\"unchen, Theresienstrasse 37, 80333 M\"unchen, Germany.}

\address{$^2$ Institut f\"ur Theoretische Physik, Universit\"at Erlangen-N\"urnberg, Staudtstra\ss e 7, 91058 Erlangen, Germany.}
\ead{neher@kitp.ucsb.edu}


\begin{abstract} 
Global physical properties of random media change qualitatively at a percolation threshold, where isolated
clusters merge to form one infinite connected component. The precise knowledge of percolation 
thresholds is thus of paramount importance. 
For two dimensional lattice graphs, we use the universal scaling form of the cluster size distributions to
derive a relation between the mean Euler characteristic 
of the critical percolation patterns and the threshold density $\pc$. 
From this relation, we deduce a simple rule to estimate $\pc$, which is 
remarkably accurate. 
We present some evidence that similar relations might hold for continuum percolation
and percolation in higher dimensions.
\end{abstract}

\maketitle

Consider a regular $d$-dimensional lattice where a fraction of sites is selected independently
with probability $p$ and deemed 'black', with the complementary vertices said to be white.  
The aggregate of these spatial lattice elements forms a  random pattern, which we may partition 
into clusters after specifying a neighborhood. 
This simple set-up constitutes the standard model of Bernoullian percolation theory, and is applied to problems as 
diverse as transport in disordered media, epidemics, and the quark confinement transition in the
early universe \citep{Isichenko_RevModPhys_92, Newman_PRE_02, Fortunato_PhysLettB_00}.
The central result of this theory is the existence of a sharp threshold value $0<\pc<1$
in an infinite lattice of dimension $d>1$: 
When $p$ increases across $\pc$, a single infinite cluster appears almost surely
and grows in mass with increasing $p$ beyond $\pc$ \citep{Hughes_96, Stauffer_95, Grimmett_99}.

Research in percolation theory focussed predominantly on the \emph{universal} critical 
phenomena showing up in the vicinity of the threshold. 
On the other hand, for practical application of percolation concepts, it is the specific and 
\emph{non-universal} value of $\pc$ which is of primary importance.
Exact values of $\pc$ are known only for special classes of 2-d lattices 
\citep{Sykes_63, Kesten_80,  Wierman_JPhysA_84, Ziff_PRE_06, Scullard_PRE_06, Wu_PRL_06}.
In all other cases, values of $\pc$ are estimated numerically with computer simulations,
which often are time consuming, in particular in three or higher dimensional lattices.

Here, we investigate the signature of the percolation transition in the 
\emph{Euler Characteristic}  of the spatial pattern formed by the percolating clusters. 
Its mean value per site, $\mecb(\pb)$, provides a topological descriptor,
which for a lattice $\lattice$ turns out to be an exactly calculable finite
polynomial in $\pb$.
For 2d-lattices the polynomials $\mecb(\pb)$ have one non-trivial zero $0<\pz(\lattice)<1$.
From the comparison with known threshold values, we observe that $\pz(\lattice)$ 
gives a tight upper bound for $\pc(\lattice)$ for many lattices, but we also find exceptions to
this rule-of-thumb.
In the case of 3d-lattices, each $\mecb(\pb)$-polynomial
has two distinct nontrivial zeros which are again slightly larger than the thresholds values
of the two distinct percolation transitions of black and white clusters.

For 2d-lattices, we explain this peculiar ordering of $\pz$ and $\pc$ using the
known scaling expression for the critical percolation clusters at $\pc$.
Moreover, this approach leads to a surprisingly simple relation which combines via $\mecb(\pb)$ the specific lattice
geometry with universal critical percolation features into an accurate
\emph{parameter-free} estimate of percolation thresholds of all 2d-lattices 
considered in this note. Our work also applies to bond percolation problems when they are reformulated as
the equivalent site percolation problem on the covering lattice.

\section{The Euler characteristic in percolation theory}
The percolation transition is a paradigm of a non-thermal phase transition,
where the local merging of black clusters causes an abrupt change in the large scale-connectivity
of black vertices.
Since the Euler Characteristic (EC) is a prominent descriptor of global aspects of spatial patterns,
we may expect it to be also a valuable tool in the study of the percolation transition. 
In this section, we introduce the EC descriptively and discuss its salient
features; a more technical but elementary outline can be found in the supplementary notes. 

\begin{figure}
\centering
\includegraphics[width=0.8\columnwidth]{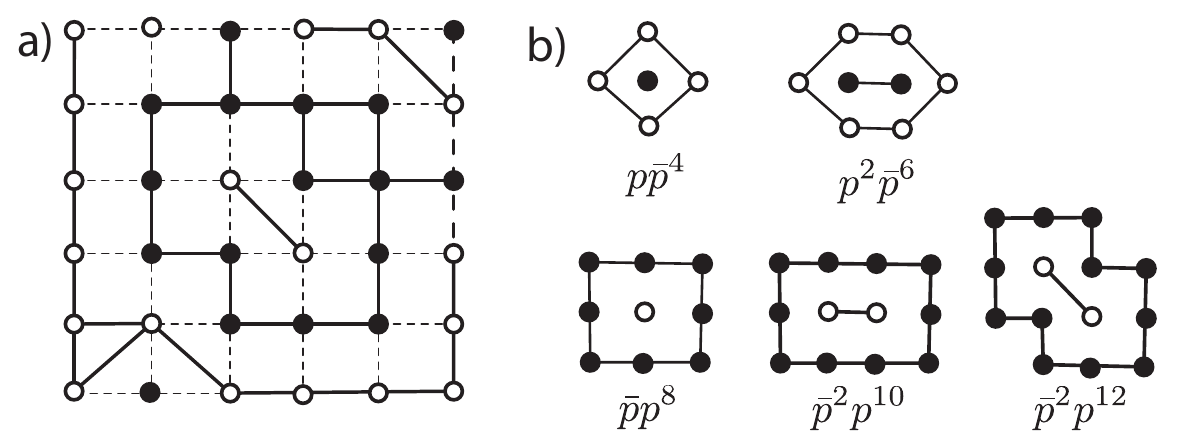}
\caption{\label{fig:lattice_clusters}
a) Black clusters partition the white vertices into an aggregate of complementary clusters. For the square lattice, 
white vertices are connected by lattice bonds \emph{and} by diagonal bonds across the faces of the lattice.
b) The black and white clusters of size 1 and 2 on the square lattice. The white perimeter
sites of a black cluster are connected (and vice versa).
}
\end{figure}

For the time being we consider planar lattices with cyclic boundary conditions. The basic object
in site percolation are clusters of vertices, naturally defined by 
the connectivity of the host lattice: Two black vertices belong to the same cluster
if they are joined by a path of black nearest-neighbors on the lattice. 
Moreover, each configuration of black clusters
specifies in a natural way an aggregate of white clusters with a complementary neighborhood,
which is in general distinct from the black one; it may be visualized by drawing ``matching"
bonds diagonally across polygonal faces of the host lattice, as illustrated in \FIG{lattice_clusters}a
for the square lattice. The lattice comprising the vertices of the original lattice and all bonds between neighboring
white vertices is called the \emph{matching lattice}.
The complementary connectivities of black and white clusters imply, that
the white perimeter vertices of black clusters form closed boundaries of holes in white clusters
and vice versa, as illustrated for black and white clusters of size one and two in \FIG{lattice_clusters}b.
The percolation thresholds of a lattice $\lattice$ and its matching lattice $\matchlattice$ add up to 
one \citep{Sykes_64}
\begin{equation}
\label{eq:matching}
\pc(\lattice)+\pc(\matchlattice)=1,
\end{equation}
implying that the black and white clusters are simultaneously critical at $\pb=\pc$.

Let $\perpolb_{st}$ denote the number of black clusters per lattice site with a fixed size $s$ and 
a fixed number $t$ of perimeter vertices. The mean density of finite black clusters
is then given by $\clusternob(\pb)=\sum_s \clusternob_s(\pb)=\sum_{st}\pb^s \perpolb_{st}\pw^t$,
where $\pw=1-\pb$ is the density of white vertices, comp. \FIG{lattice_clusters}b. Correspondingly, the mean density
of the complementary finite white clusters reads 
$\clusternow(\pw)=\sum_s \clusternow_s(\pw)=\sum_{st}\pw^s \perpolw_{st}\pb^t$.
In their pioneering paper \citep{Sykes_64} on exact percolation thresholds in two dimensions,
\citeauthor{Sykes_64} considered the difference 
\begin{equation}
\label{eq:ec_series}
\mecb(\pb):=\clusternob(\pb)-\clusternow(\pw)=-\mecw(\pw)
\end{equation}
and found that $\mecb(\pb)$ is a finite polynomial, which they called \emph{matching polynomial}.
This observation enabled them to obtain, for instance, the exact critical probability of site percolation
on the ``self-matching" triangular lattice from $\mecb_{tri}(\pc)=0$ at $\pc=\frac{1}{2}$.

\begin{figure}
\centering
\includegraphics[width=0.8\columnwidth]{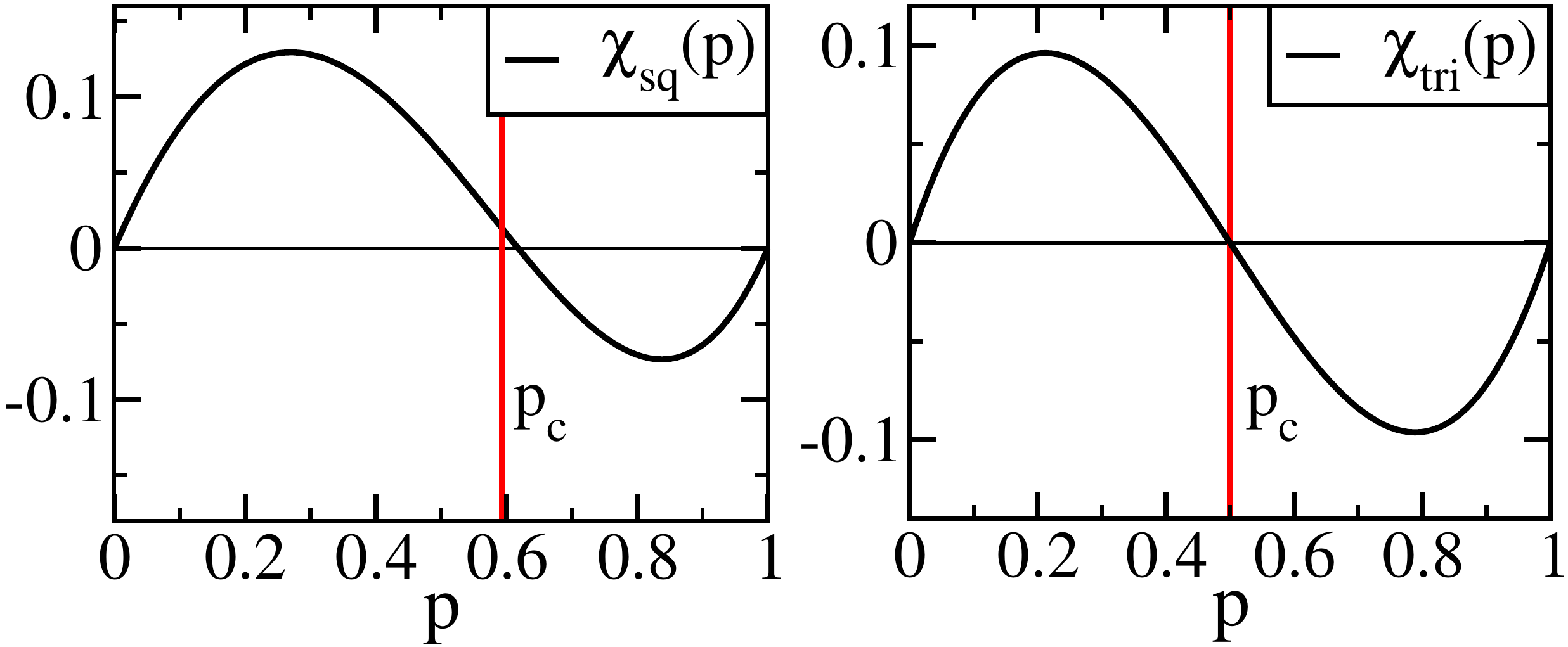}
\caption{\label{fig:mec_square_cubic}The MEC of the square lattice (left) and of the triangular lattice (right).
Note, that $\pz$ is slightly above $\pc$ for the square lattice. The triangular lattice is self-matching and $\pc=\pz=1/2$. }
\end{figure}

When specialized to spatial patterns $\patternb_N$ on a planar lattice with $N$ sites,
the definition of the Euler characteristic, $\EC$, reads
\begin{equation}
\label{eq:ec_clusters_holes}
\EC(\patternb_N)=\#Ê\mathrm{\;clusters\;of\;} \patternb_N-\#Ê\mathrm{\;holes\;in\;} \patternb_N.
\end{equation}
From comparison of this definition with \EQ{ec_series}, and the observation that complementary 
(finite) white clusters constitute the holes in black clusters (see above), we see that the matching polynomial $\mecb(\pb)$
may be identified with the mean Euler characteristic per site (MEC) of the black clusters, 
$\mecb(\pb)=\lim_{N\to\infty}\langle\frac{1}{N}\EC(\patternb_N)\rangle_\pb$.
\EQ{ec_series} expresses the fundamental topological invariance of the EC,
but the representation as the difference of two infinite series is not convenient for practical computation. 
For that purpose we employ Euler's polyhedral formula, which expresses the EC
\begin{equation}
\label{eq:ec_polyhedral}
\EC=\#v- \#e+\#f
\end{equation}
in terms of the number of black vertices $\#v$, edges joining black vertices $\#e$ and polygonal faces
with black boundary $\#f$, respectively. The mean value $\mecb(\pb)$ is now obtained by a simple local 
calculation. As an example, consider the square lattice: 
(i) A vertex is  black with probability $\pb$.
(ii) The two vertices bounding an edge are black with probability $p^2$, and there are two edges per vertex.
(ii) The four vertices surrounding a face are black with probability $\pb^4$.
Hence, we find for the square lattice
\begin{equation}
\label{eq:ec_square}
\mecb_{sq}(\pb)=\pb -2\pb^2+ \pb^4. 
\end{equation}
The graph of $\mecb_{sq}(\pb)$ is shown in \FIG{mec_square_cubic}a.
Analogously, one finds for the triangular lattice $\mecb_{tri}(\pb)=\pb -3\pb^2+ 2\pb^3$, which has the
above mentioned self-matching property $\mecb(\pb)=-\mecw(1-\pb)$ (see \FIG{mec_square_cubic}b).
Whenever the lattices cells have finite number of boundary vertices,  $\mecb(\pb)$ is a finite polynomial.

The graph of $\mecb_{sq}(\pb)$ in  \FIG{mec_square_cubic} is typical for the MEC of 2d-lattices. At small values of  $\pb$, black 
clusters are \emph{finite} holes in a \emph{single infinite white} cluster. As long as $\pb$ is well below $\pc$, the 
density of holes within the small-sized black clusters is negligible, hence $\mecb(\pb)$ is positive and 
increases with increasing $\pb$.
On the other hand, for $\pb > \pc$,  and $1-\pb \ll 1$, there is a \emph{single infinite black} cluster with 
\emph{finite} (white) holes and thus $\mecb(\pb)<0$, in accordance with \EQ{ec_clusters_holes}. 
In the intermediate range of $\pb$, which includes $\pc$, the MEC decreases as the black clusters 
grow in size and merge to generate a single infinite component as $\pb$ passes the percolation 
threshold at $\pc$. 
We see that the typical features of the MEC are governed by the interplay of the complementary 
finite black and white clusters. The percolation transition with the singular emergence of an 
infinite cluster leaves its signature only in the zero crossing of  $\mecb(\pb)$ at $0 < \pz  < 1$ with a 
value of $\pz$ expected to be comparable with $\pc$. 

\section{Percolation thresholds and the zero crossing of the MEC}
The idea that the zero crossing of the MEC should occur near the critical probability $\pc$ is 
plausible and is supported, for instance, by the fact that $\pz = \pc =1/2$ for site percolation on the 
self-matching triangular lattice, but it calls for a more precise and quantitative argument. 
Here we will first compare $\pz$ with $\pc$ and we shall find that $\pz$ provides generally 
a tight upper bound to $\pc$ whenever $\pc > 1/2$.
\begin{figure}
\centering
\includegraphics[width=0.45\columnwidth]{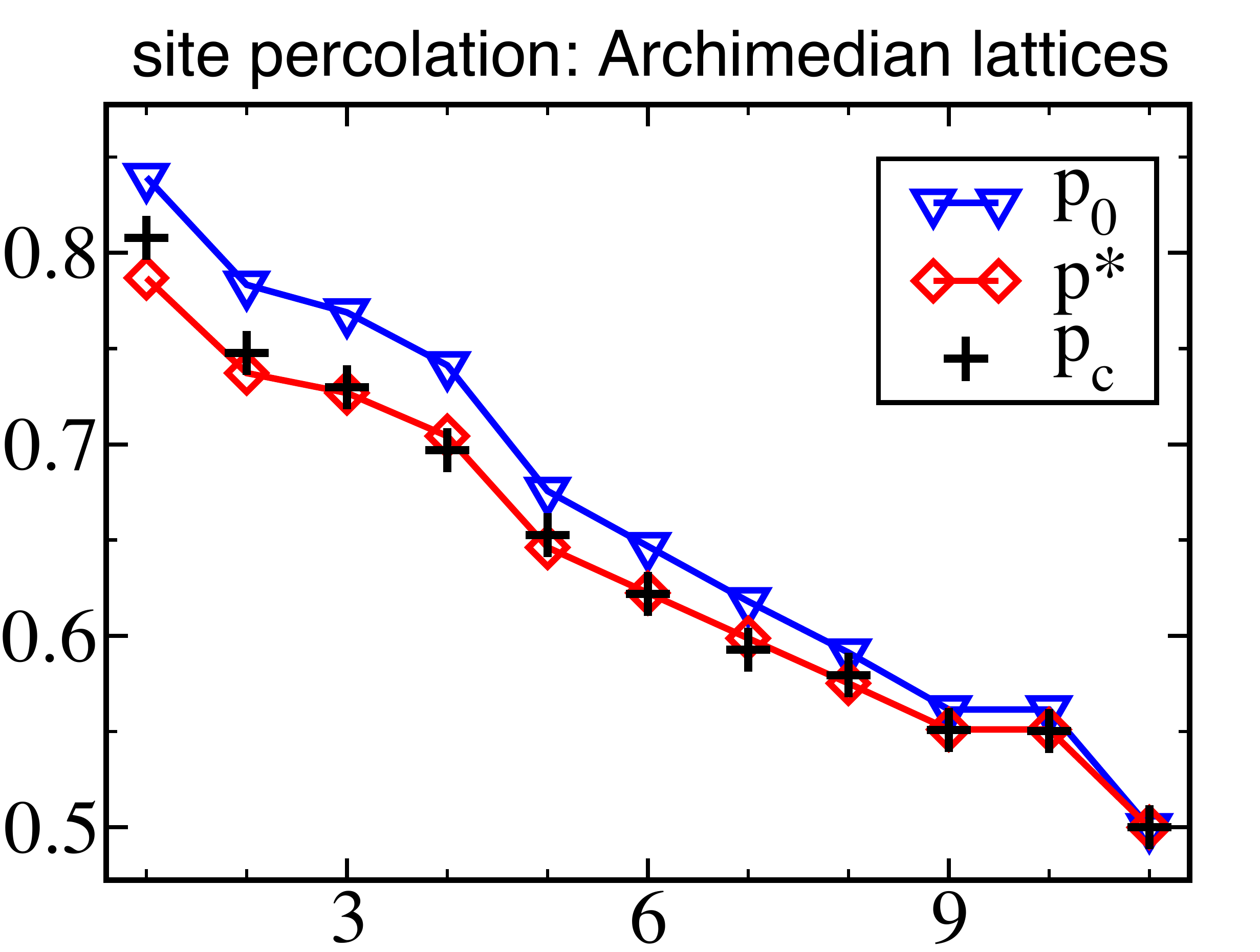}
\includegraphics[width=0.45\columnwidth]{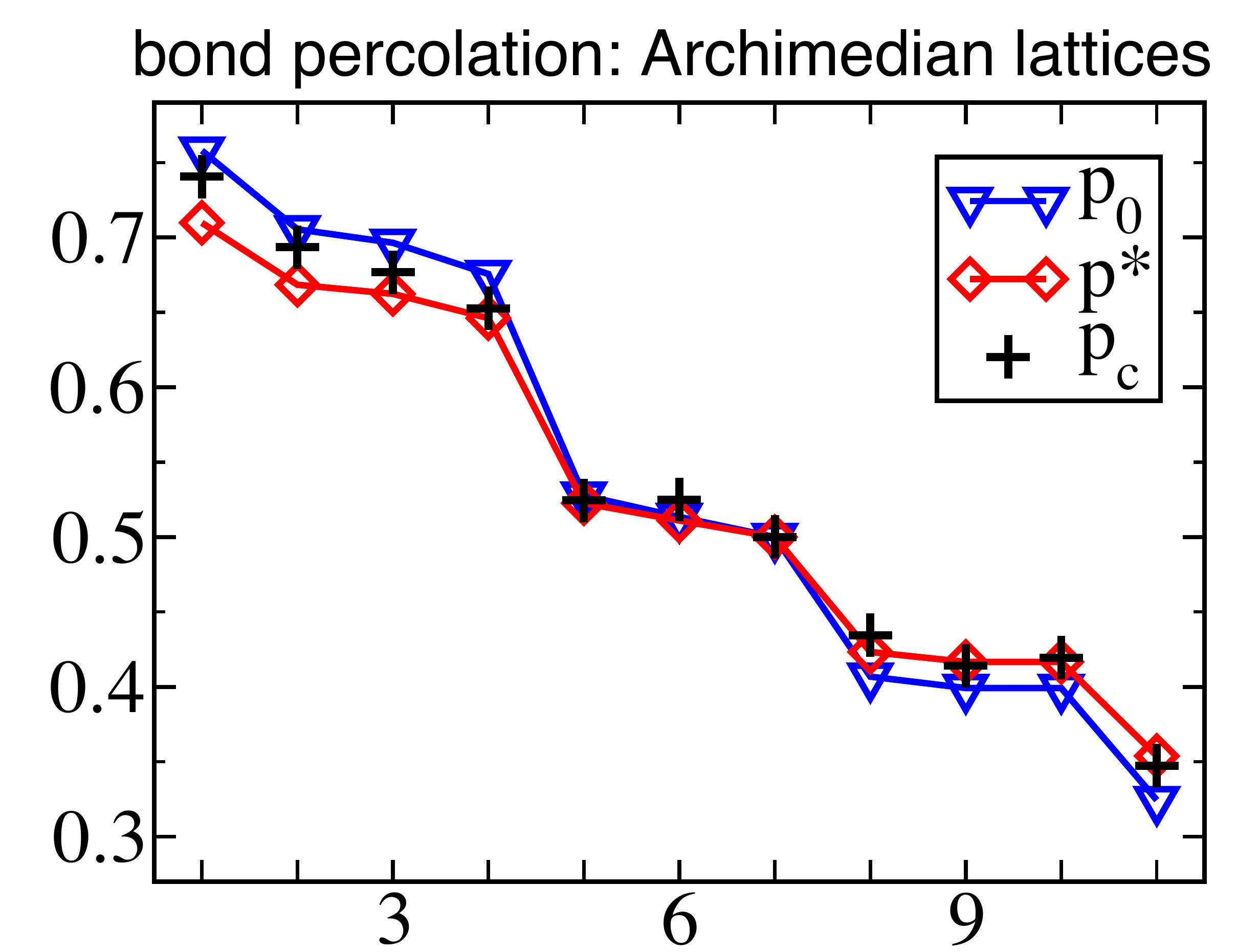}
\includegraphics[width=0.52\columnwidth]{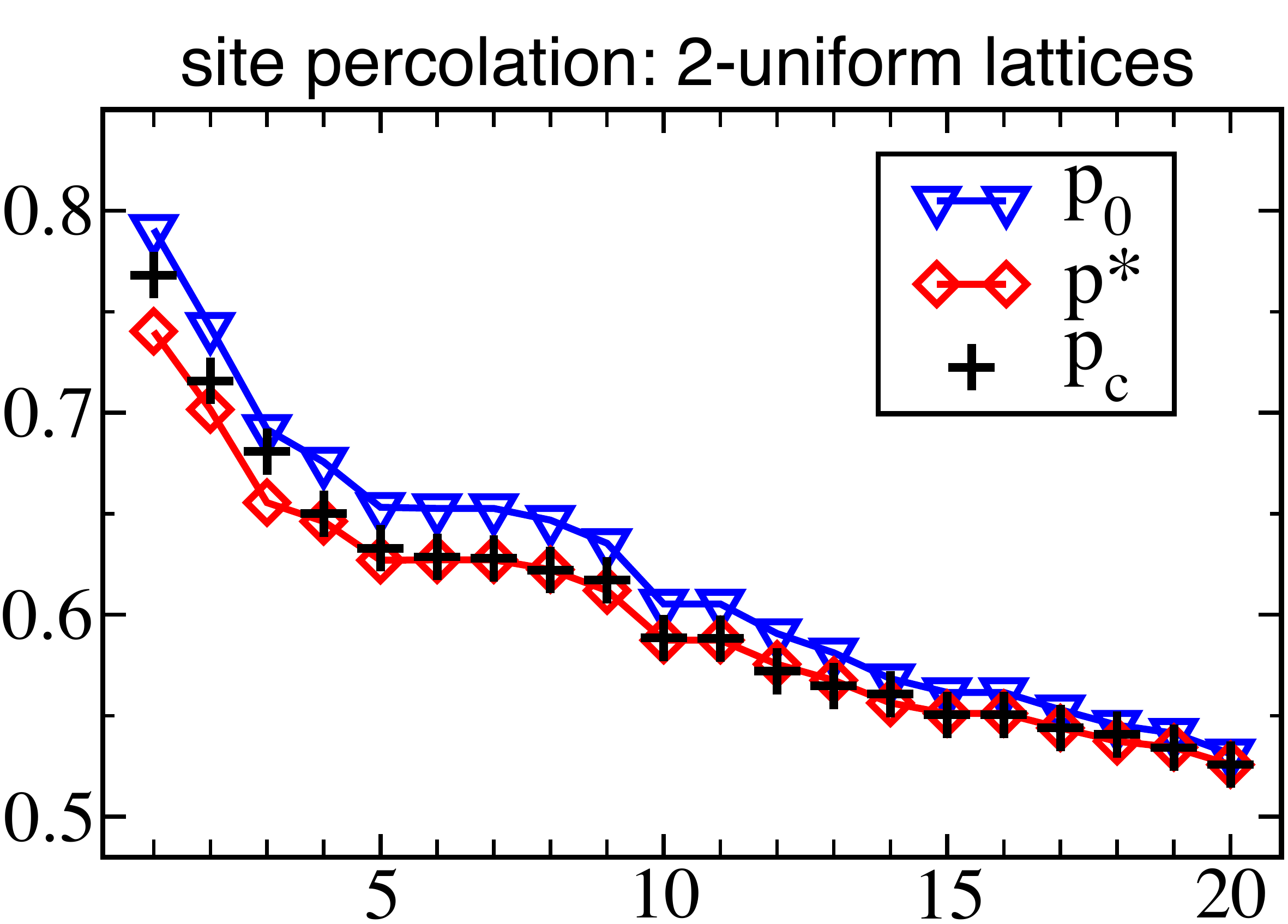}
\caption{\label{fig:lattices} The percolation threshold $\pc$ is slightly below the zero crossing
$\pz$ of the MEC whenever $\pc>\frac{1}{2}$. This order is reversed, if $\pc<\frac{1}{2}$, as apparent 
in the panel on bond percolation. The solution $\pest$ of \EQ{pc_estimation} provides 
a very accurate estimate of $\pc$. The deviation $|\pc-\pest|$ exceeds 0.01 only for very open lattices
with high percolation thresholds. Lattices are in the order of decreasing percolation threshold. 
For vertex configurations, numerical values of $\pc$
and $\pest$ for the 2-uniform lattices, see Tab.~1 in the supplementary material.
Numerical estimates of $\pc^{site}$ for the Archimedean lattices are taken from \citep{Suding_PRE_99},
values for $\pc^{bond}$ are from \citep{Parviainen_JPA_07, Marck_03}.
}
\end{figure}

\subsection{Archimedean lattices } 
To begin with, we consider the eleven Archimedean lattices, where 
all vertices are equivalent up to a symmetry operation and all faces are regular polygons. The 
most prominent members of this class are the triagonal, the square, the honeycomb and the kagom\'e lattice.
Each lattice is uniquely characterized by the number of edges $n_i$ of the polygons 
surrounding a vertex \citep{Gruenbaum_86}. A lattice of coordination number $z$
is therefore conveniently denoted by the symbols $(n_1, \ldots, n_z)$, where $a_i$ identical consecutive polygons 
are often abbreviated as $n_i^{a_i}$. For example, a vertex of the square lattice is surrounded by 
four squares. The vertex type of the square lattice is therefore $(4,4,4,4)$ or $(4^4)$ in the abbreviated notation. 
Similarly, the a vertex of the kagom\'e lattice is surrounded by alternating triangles and hexagons and has
vertex type $(3,6,3,6)$. With this notation, the MECs of the Archimedean lattices are given by
\begin{equation}
\label{eq:ec_archimed}
 \mecb(\pb)=\pb(1-\pb)\left( 1-\pb\sum_{i=1}^{z}\frac{1}{n_{i}}\sum_{\mu=0}^{n_{i}-3}\pb^\mu\right).
\end{equation}
The percolation thresholds  of Archimedean lattices are known
to very high precision \citep{Suding_PRE_99}. In \FIG{lattices}, we compare $\pc$ to the 
zero crossing $\pz$ of $\mecb(p)$. For the self-matching triangular lattice $\pz=\pc=\frac{1}{2}$.
Furthermore, a close relation appears to exist between $\pc$ and $\pz$ even for lattices with $\pc>\frac{1}{2}$:
$\pc$ is bounded from above by $\pz$, with $\pz-\pc$ increasing 
steadily with $\pc-\frac{1}{2}$. Similar observations can be made for the duals of the Archimedean lattices, 
see Fig.~2 of the supplementary material.
 
\subsection{2-uniform lattices} 
Next, we study the larger class of 2-uniform lattices, which again consist of regular polygons
but have two distinct vertex types \citep{Gruenbaum_86}. 
The two vertex types can occur with different abundances, and a 2-uniform lattice is commonly 
denoted by $s_1(n_1^1, \ldots, n_{z_1}^1)+s_2(n_1^2, \ldots, n_{z_2}^2)$, where $s_i$ is the fraction of
vertices that are of type $i$. The MECs are given by the straightforward generalization of \EQ{ec_archimed}
\begin{equation}
\label{eq:ec_2uniform}
  \mecb(\pb)=\pb(1-\pb)\left( 1-\pb\sum_{\nu=1,2}\sum_{i=1}^{z_\nu}\frac{s_\nu}{n_{i}^\nu}\sum_{\mu=0}^{n_{i}^\nu-3}\pb^\mu \right),
\end{equation}
which can obviously be generalized to any finite number of vertex types. 
We are not aware that the percolation thresholds of 2-uniform lattices have been previously determined,
and we therefore estimated them using an algorithm adopted from
\citep{Newman_PRE_01}. Again, $\pz$ provides a tight upper bound to $\pc$ as shown in \FIG{lattices}.
The vertex configurations, as well as $\pz$  and the estimate of $\pc$ for the 2-uniform lattices are given 
in Table 4 of the supplementary material.

\subsection{Bond percolation in two dimensions}
Every bond percolation problem is equivalent to site percolation on the covering lattice. The covering
lattices, however, are not necessarily planar but decorated mosaics. 
A decorated mosaic is constructed from a planar lattice, where in a subset of the faces all diagonal
connection have been added (the face is decorated). A pair of lattices where complementary 
sets of faces are decorated constitute a pair of matching lattices in the sense of \EQ{matching} \citep{Sykes_64}. 
Calculating the MEC of decorated mosaics is slightly more laborious, but 
a general framework for the calculation has been presented in \REF{Essam_JMathPhys_79}.
Using this framework, we calculated the MEC of the covering lattices of all 
Archimedean lattices with vertex $(n_1, \ldots, n_z)$
\begin{equation}
\label{eq:ec_bond}
\mecb(\pb)=-\pb+\frac{2}{z}\left(1-(1-\pb)^z\right)+\sum_i\frac{2}{zn_i}\pb^{n_i}.
\end{equation}

Comparing numerical estimates of $\pc$ \citep{Parviainen_JPA_07, Marck_03} to $\pz$ confirms
$\pz>\pc$ if $\pc>\frac{1}{2}$ (\FIG{lattices}), albeit with one notable exception for the lattice 6 with vertex configuration $(3,4,6,4)$. 
For lattices with $\pc<\frac{1}{2}$ the order of $\pc$ and $\pz$ is reversed,
as expected from the matching properties of $\pc$ and $\mec(p)$.

The relation between $\pc$ and $\pz$ is not restricted to regular lattices but also holds for 
the quasi-periodic Penrose tiling and random tessellations of the plane such as Voronoi and 
Delauny tessellations, see supplementary material for values of $\pc$ and $\pz$ and the polynomials 
of the MECs.

\subsection{Randomly decorated mosaics}
\begin{figure}
  \centering
  \includegraphics[width=0.9\columnwidth]{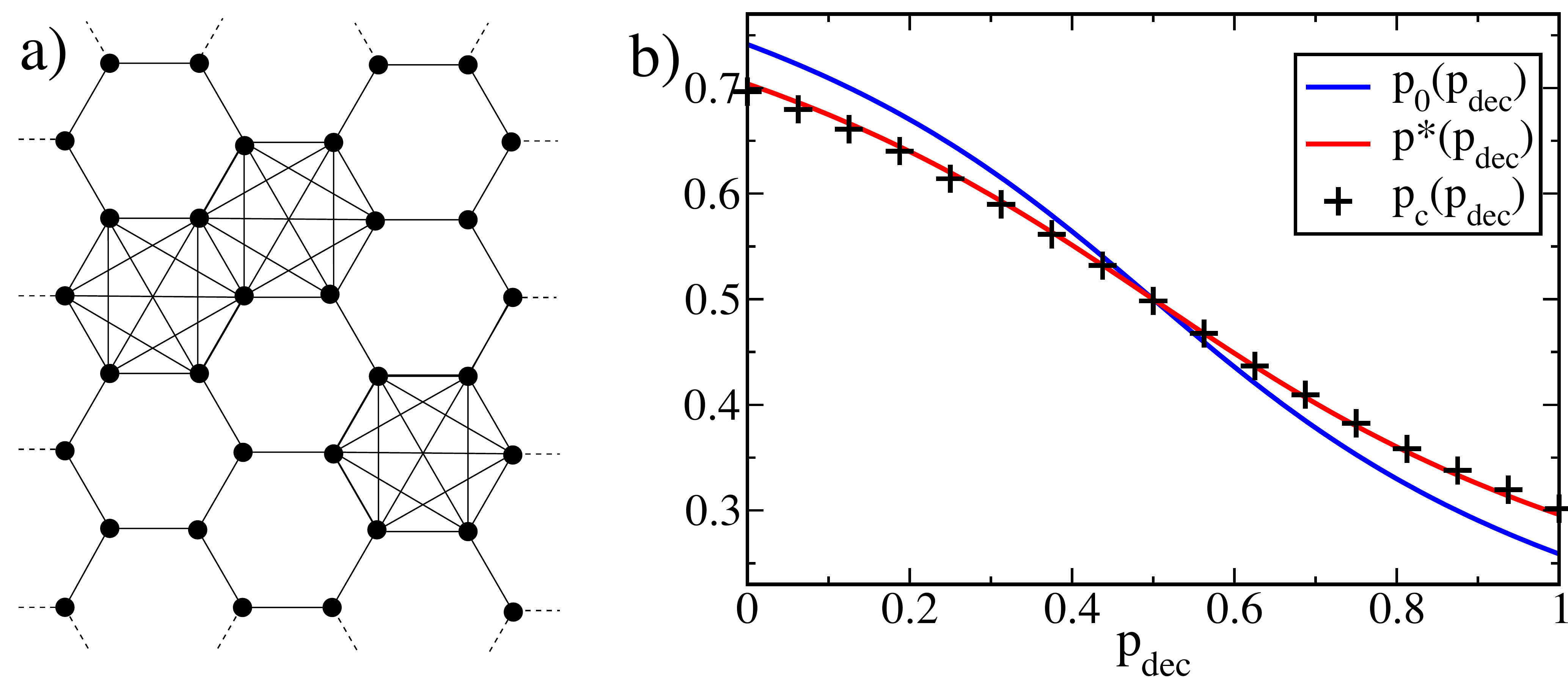}
  \caption{\label{fig:random_decoration} a) A lattice face is decorated, when all diagonal connections across the faces have been added to the lattice graph, as illustrated here for three faces of the hexagonal lattice. We consider randomly decorated lattices where each face is decorated with probability $\pdec$. b) The site percolation threshold $\pc(\pdec)$ decreases smoothly  as the degree of decoration is varied from $\pdec=0$ to $\pdec=1$. The zero crossing $\pz(\pdec)$ of $\mecb(\pb,\pdec)$ provides a tight upper to $\pc(\pdec)$ if $\pc(\pdec)>\frac{1}{2}$ and vice versa. The solution of \EQ{pc_estimation}, $\pest(\pdec)$, lies within $0.005$ of $\pc(\pdec)$ with the largest deviations at
full or no decoration. }
\end{figure}

Instead of regular decorated mosaics, we now consider lattices where each face is decorated
with probability $\pdec$, as illustrated in \FIG{random_decoration}a. We are not aware that this type of percolation process,
which bears some similarity to a bond-site percolation processes, has been studied before. 
Our numerical estimates of percolation thresholds of a randomly decorated 
mosaic decreases smoothly from $\pc(\pdec=0)$ to $\pc(\pdec=1)$, in accord with the containment property \citep{Fisher_JMathPhys_61}.
Randomly decorated lattices fulfill the statistical matching property
$\pc(\pdec)+\pc(1-\pdec)=1$, from which $\pc(0.5)=0.5$ follows. 
The averaging over the different decoration states of the lattice is straightforward and the MEC of randomly decorated
mosaics can be calculated in the same way as that of regular decorated mosaics. For the hexagonal lattice
one finds
\begin{equation}
\label{eq:ec_randomdecor}
\mecb(\pb,\pdec)=(1-\pdec)\mecb_{hex}(\pb)+\pdec\mecw_{hex}(\pb).
\end{equation}
From \EQ{ec_series} follows the symmetry relation $\mecb(\pb,\pdec)\!=\!-\!\mecb(1\!-\!\pb,\! 1\!-\!\pdec)$.
Hence, we have $\pz(\pdec)\!+\!\pz(1\!-\!\pdec)=1$ in analogy to $\pc(\pdec)\!+\!\pc(1\!-\!\pdec)\!=\!1$.
Our results for the randomly decorated hexagonal lattice are shown in \FIG{random_decoration}b.
The zero crossing $\pz(\pdec)$ follows $\pc(\pdec)$ very 
closely, being a tight upper bound for $\pdec>\frac{1}{2}$ and a lower bound otherwise. Similar results can be obtained
for other Archimedean lattices (data not shown).

\section{The EC of critical percolation and estimation of $\pc(\lattice)$}
In the previous sections, we saw that $\pc(\lattice)$ -- a global property of the lattice --
is followed rather closely by $\pz(\lattice)$, a locally computable quantity. 
Here, we are going to explore the relation of $\pc(\lattice)$ with $\pz(\lattice)$ in more detail
by evoking a generally accepted scaling form for the densities $\clusternob_s(\pb)$ of large 
clusters at $\pc$ \citep{Stauffer_PhysRep_79}.
These densities determine $\mecb(\pb)$ according to \EQ{ec_series}. Moreover, 
they also enter in the sum rules \citep{Hughes_96}
\begin{equation}
\nonumber
\pb=\sum_{s} s\clusternob_s(\pb)+\pbinf(p),\quad \pw=\sum_{s} s\clusternow_s(\pw)+\pwinf(\pw),
\end{equation}
which express the probability for a particular vertex to be black ($\pb$) or white ($\pw$).

In 2d-lattice graphs only a single critical point exists, so that $\pbinf(\pc)=0=\pwinf(\pw_c)$.
At the threshold the scaling ansatz reads
\begin{equation}
\label{eq:asymp_clustersize}
\clusternob_s(\pc)\simeq \prefacb(\lattice) s^{-\tau} \quad \mathrm{and}\quad \clusternow_s(\pw_c)\simeq \prefacw(\matchlattice) s^{-\tau}.
\end{equation}
The non-universal amplitudes $\prefacb(\lattice)$ and $\prefacw(\matchlattice)$ account for the 
particular structure of the underlying lattice and its matching partner, whereas the value
of the universal exponent $\tau=187/91$ is known exactly in two dimensions \citep{Nienhuis_JPhysA_80, Hughes_96, Smirnov_MRL_01}.

In order to exploit the scaling hypothesis, we define
\begin{equation}
\label{eq:coarse_ec}
\mec_{s_0}(\pb):=\mec(\pb)-\sum_{s=1}^{s_0}\left[\clusternob_s(\pb)-\clusternow_s(\pw)\right],
\end{equation}
and set
\begin{equation}
\label{eq:scaling_ec}
\mec_{s_0}(\pc)\simeq (\prefacb-\prefacw)\sum_{s>s_0}s^{-\tau}.
\end{equation}
Likewise, 
\begin{eqnarray}
\label{eq:coarse_density}
\dpq_{s_0}(\pb)&:=&(\pb-\pw)-\sum_{s=1}^{s_0}s\left[\clusternob_s(\pb)-\clusternow_s(\pw)\right];\\
\label{eq:scaling_density}
\dpq_{s_0}(\pc)&\simeq&(\prefacb-\prefacw)\sum_{s>s_0}s^{1-\tau}.
\end{eqnarray}
After elimination of the non-universal scaling amplitudes, we arrive at
\begin{equation}
\label{eq:scaling_arg}
\mec_{s_0}(\pc)\simeq\frac{\zeta(\tau, s_0)}{\zeta(\tau-1, s_0)}\dpq_{s_0}(\pc), 
\end{equation}
where $\zeta(\tau, s_0)=\sum_{s=1}^{\infty}(s+s_0)^{-\tau}$ is the Riemann Zeta-function with 
offset $s_0$.

\begin{figure}
  \centering
  \includegraphics[width=0.6\columnwidth]{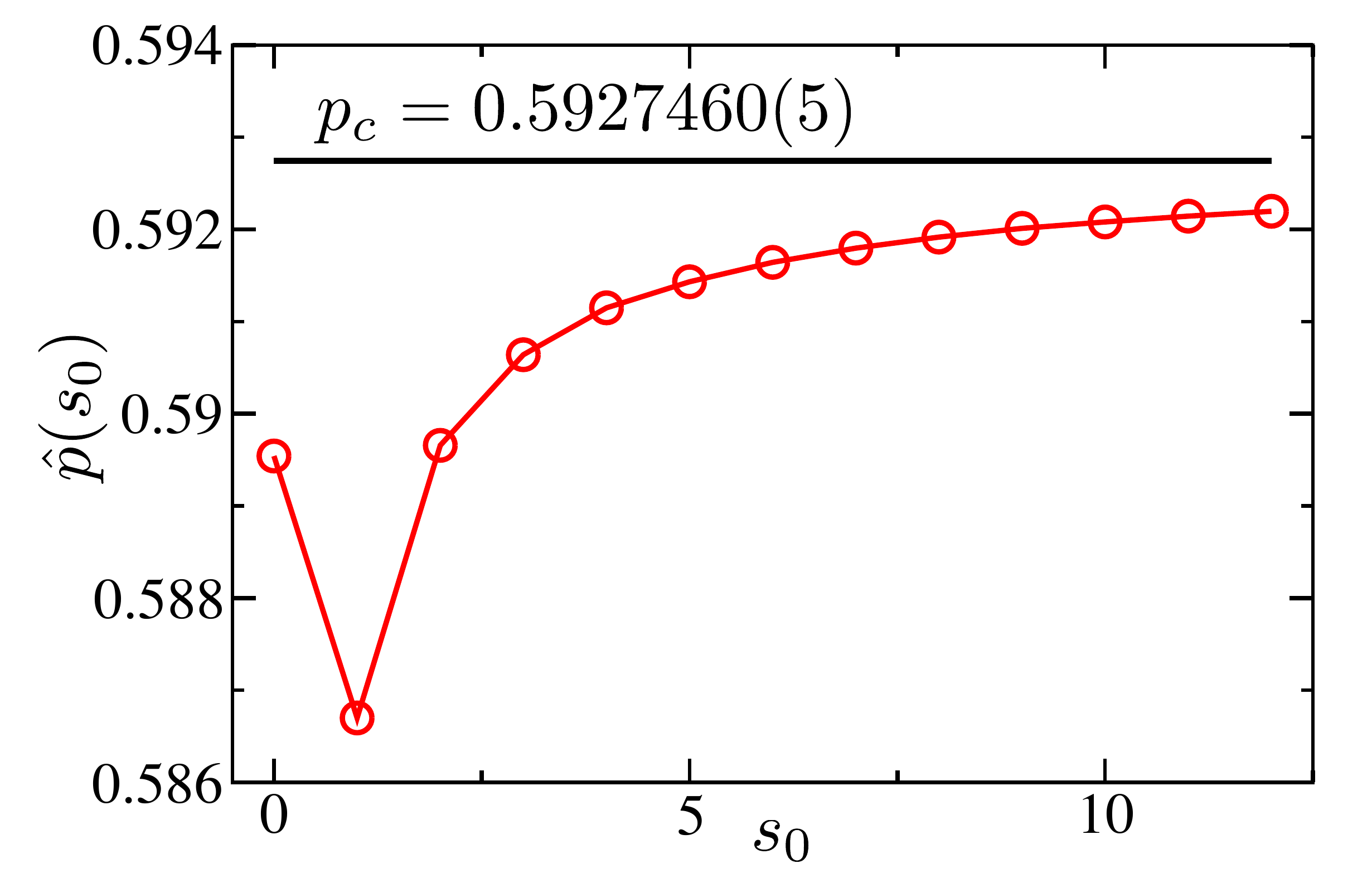}
  \caption{\label{fig:estimation} Square lattice -- site percolation: The solution $\pestzeta(s_0)$ of \EQ{ec_crit_percolation}  
approaches $\pc$ with increasing $s_0$. 
}
\end{figure}

The relation (\ref{eq:scaling_arg}) may be applied as an equality, for instance,
(i) to determine a value of $s_0$ from the requirement that the left- and right-hand sides
equalize within a prescribed accuracy, or 
(ii) to estimate $\pc(\lattice)$.
For the latter purpose, we rewrite \EQ{scaling_arg} by substituting the defining expressions (\ref{eq:scaling_ec}, \ref{eq:scaling_density}) for $\mec_{s_0}(\pb)$ and $\dpq_{s_0}(\pb)$. The result is
\begin{eqnarray}
\label{eq:ec_crit_percolation}
\mecb(\pb)=&\frac{\zeta(\tau, s_0)}{\zeta(\tau-1, s_0)}\left[\left( 2\pb -1\right)-\sum_{s=1}^{s_0}s\left[\clusternob_s(\pb)-\clusternow_s(\pw)\right]\right]\\\nonumber
&+\sum_{s=1}^{s_0}\left[\clusternob_s(\pb)-\clusternow_s(\pw)\right].
\end{eqnarray}
The real root, $\pestzeta(s_0)$, $0<\pestzeta(s_0)<1$, of this polynomial equation provides an estimate for $\pc$, the accuracy of which increases
with $s_0$. For the square lattice,
the cluster numbers $\perpolb_{st}$ and $\perpolw_{st}$ are known up to $s=12$ \citep{Mertens_90} and we calculated $\pestzeta(s_0)$
for $s_0=0,\ldots,12$. \FIG{estimation} shows how $\pestzeta(s_0)$ approaches $\pc$ with increasing $s_0$. 
The Zeta functions $\zeta(\tau,s_0)$ and $\zeta(\tau-1,s_0)$ can be approximated by integrals 
and their ratio evaluates to $(\tau-2)(\tau-1)^{-1}(s_0+1)^{-1}$. 
In many cases corrections for small clusters are not even necessary, and using $s_0=0$ and the 
approximation for the Zeta function yields the simple equation 
\begin{equation}
\label{eq:pc_estimation}
\mec(\pb)=\frac{\tau-2}{\tau-1}(2\pb-1),
\end{equation}
with a unique solution $\pest$, $0<\pest<1$. 
For all lattices discussed so far, $\pest$ is a fairly accurate estimate of $\pc$,
as shown in the Figures \ref{fig:lattices} and \ref{fig:random_decoration}\footnote{In many cases, the integral approximation
of the Zeta function yields better results than the exact ratio, which is probably due to a subtle cancelation of 
errors.}. Further examples may be found in the supplementary note. 
Among the lattices studied here, the deviation of $\pest$ from $\pc$ was greatest for
the lattices with very high percolation thresholds (comp.~\FIG{lattices}a and \FIG{lattices}b)
which are inhomogeneous on small scales. To test whether these deviations are caused by the 
smallest clusters and holes, we enumerated the clusters and holes of size one and two for the hexagonal lattice,
the $(4,8^2)$ lattice, the $(4,6,12)$ lattice, the $(3,12^{2})$ lattice, and the 2-uniform lattice 
with vertex configuration $\frac{1}{2}(3,4,3,12)+\frac{1}{2}(3,12^{2})$. The deviation
of $\pestzeta(s_0)$ from $\pc$ decreases when the contributions of the smallest clusters and holes are 
subtracted, i.e. $s_0$ is increased from 0 to 2, see \TAB{small_clusters}.

The estimation of the threshold value $\pc$ via \EQ{pc_estimation} will fail, if small clusters or 
small holes are much more abundant at the critical point than an extrapolation of the asymptotic law
for large clusters would suggest. In particular, this is obvious for lattices that contain a substructure which
does not contribute to large scale connectivity but dominates the density of small clusters.

\begin{table}
\caption{\label{tab:small_clusters} The deviation of the estimate $\pestzeta(s_0,\lattice)$ from $\pc(\lattice)$ decreases rapidly when $s_0$ is increased from zero to two.}
\begin{indented}
\item[]\begin{tabular}{@{} r r r r r r }
\br
$s_0$ & $6^3$ & $4,8^{2}$& $4,6,12$&$3,12^{2}$& $\frac{1}{2}  {{3,4,3,12}\choose{3,12^{2}}}$ \cr
\hline
0&-0.01027&-0.02258&-0.03184&-0.04543&-0.05085\cr
1&-0.00553&-0.00901&-0.01530&-0.01240&-0.02093\cr
2&-0.00551&-0.00283&-0.00701&-0.00615&-0.00937\cr
\hline
\end{tabular}
\end{indented}
\end{table}

\section{Three dimensional lattices}
The combinatorial EC of percolation patterns $\patternb$ on a $d$-dimensional lattice is given 
by the alternating sum of the numbers of $k$-dimensional lattice cells, $k = 0,\ldots,d$, 
contained in $\patternb$. Thus, in the case $d = 3$, the \EQ{ec_polyhedral} is replaced by 
\begin{equation}
\label{eq:ec_threed}
\EC(\patternb) = \#v-\#e+\#f-\#c, 
\end{equation}
where $\#c$ is the number of black three-dimensional polyhedral lattice cells. 
The resulting MEC of black clusters in the case of site-percolation on a simple 
cubic (sc) lattice, for example, is given by 
\begin{equation}
\mecb_{sc}(\pb) = \pb-3\pb^2+3\pb^4-\pb^8.
\end{equation}
The black vertices are connected, i.e. they are part of the same cluster, if they are joined by a
path of black nearest neighbors on the lattice. As in the two dimensional case, the black 
clusters partition the white vertices into an aggregate of clusters with a complementary
connectivity, such
that there is a one-to-one correspondence between cavities in black clusters
and finite white clusters, as well as between the cavities in white clusters
and finite black clusters. 
A white vertex of the simple cubic lattice, for example, is connected to all 26 vertices
on the boundary of the eight surrounding lattice cubes, in contrast to the six 
neighbors of the black vertices, which correspond to lattice bonds.
In addition to cavities, black clusters can have handles, i.e. 
they can be homeomorph to a solid torus or objects of higher genus. Each handle of a black cluster
is pierced by precisely one handle of a white cluster. 
From these duality relations and the 3d analogue of \EQ{ec_clusters_holes}
\begin{equation}
\EC(\patternb) = \#\mathrm{clusters}-\#\mathrm{handles}+\#\mathrm{cavities},
\end{equation}
we see that EC of white clusters is identical to that of black clusters. Hence, 
MEC of white clusters is $\mecw(\pw)=\mecb(1-\pw)$ with $\pw=1-\pb$.

\begin{figure}
\centering
\includegraphics[width=0.8\columnwidth]{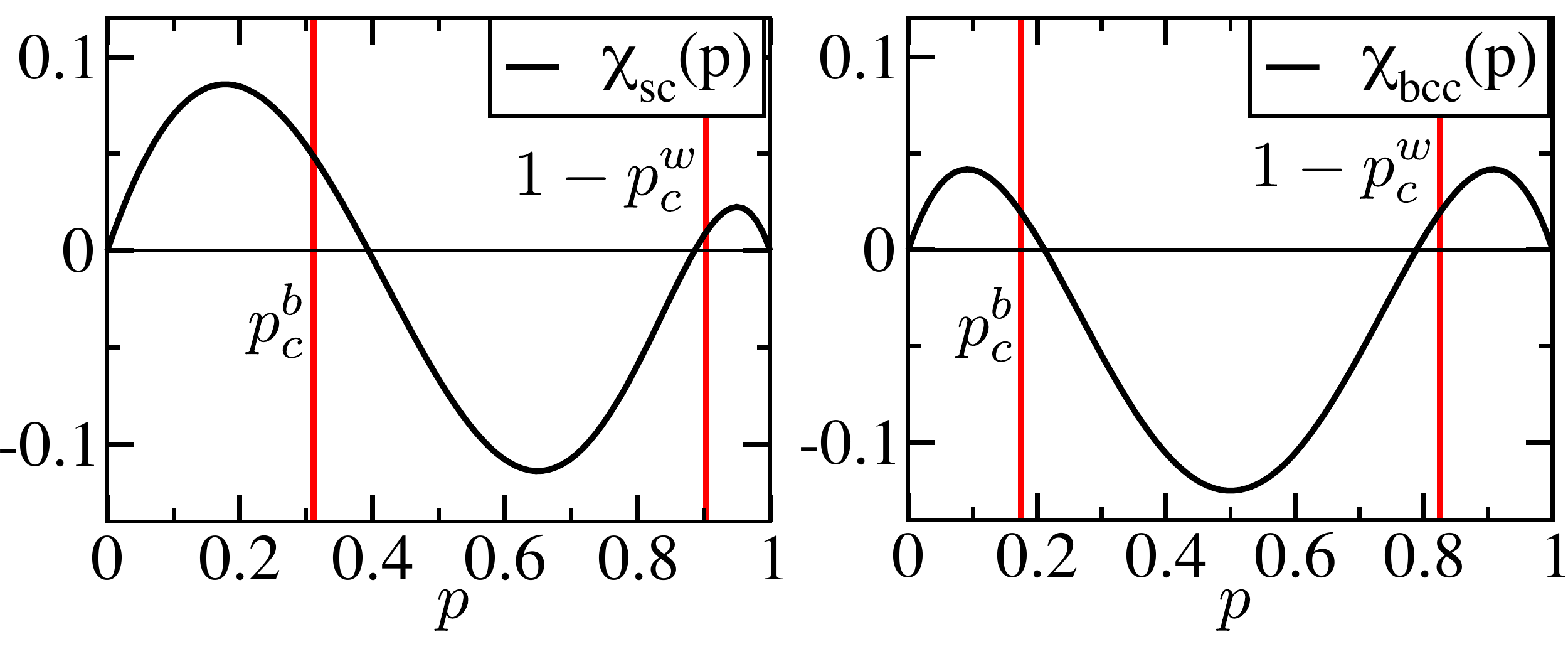}
\caption{\label{fig:mec_cubic}The graphs of the MEC of the simple cubic (sc) and body-centered cubic (bcc) lattice.
The bcc lattice has the same black and white connectivities and is hence invariant to the 
substitution $\pb=1-\pb$.}
\end{figure}

The graph of $\mecb_{sc}(p)$ shown in \FIG{mec_cubic} is typical for the MECs in $d=3$,
where the MECs have two distinct non-trivial zero crossings $\pbz$ and $1-\pwz$.
In the intermediate regime where $\mecb(\pb)<0$, the MEC is dominated by interwoven
white and black handles.
The percolation thresholds of the lattice with black  $\pbc$ and white $\pwc$ connectivity
are different in general.  
In the range $\pbc < \pb < 1-\pwc$ a single infinite black cluster coexists with a single 
infinite white cluster. 
In order to check for a possible link between zero crossings of $\mecb(\pb)$ and thresholds 
we compare $\pbz$ and $\pwz$ with simulation values $\pbc$ and $\pwc$ for the sc lattice, face-centered-cubic (fcc)
lattice and the body-centered-cubic (bcc) lattice. The calculation of the MECs for 
fcc and bcc lattices is reported in the supplementary note.
As the \FIG{mec_cubic} already indicates, $\pbz$ and $\pwz$ are both upper bounds 
for $\pbc$ and $\pwc$, and they are becoming tighter as the (effective) coordination 
numbers increase. 
The task to device a threshold estimator based on the above findings appears to 
be more difficult than in the two dimensional case, and it is left for future work.

\begin{table}
\caption{\label{tab:cubic_lattices} Numerical values for $\pz$ and $\pc$ of the cubic lattices for black and white connectivities. The bcc lattice has equal black and white connectivities. Threshold
densities are taken from ${}^a$)\citep{Lorenz_JPhysA_98}, ${}^b)$\citep{Essam_72}}
\begin{indented}
\item[]\begin{tabular}{@{}lrrr}
\br
lattice& $\quad\quad z$ & $\quad\quad\quad\pz$ &$\quad\quad\quad\pc$\\
\hline
\hline
sc (black) & 6 &0.3940& 0.3116${}^a$\cr
fcc (black)& 12&0.2370 & 0.1992${}^a$\cr
bcc& 14&0.2113 & 0.175${}^b$\cr
fcc (white)& 18&0.1616 & 0.136${}^b$\cr
sc (white)& 26 &0.1139& 0.097${}^b$\cr
\hline
\end{tabular}
\end{indented}
\end{table}

\section{Continuum percolation}
So far, we dealt with the EC of percolating clusters on geometric lattices. Let us 
finally make a few remarks to indicate that the features of the EC induced by the 
percolation thresholds persist in the case of continuum percolation. 
Consider the standard Boolean model where penetrable convex grains are
positioned randomly at Poisson distributed points in $\mathbb{R}^d$. The grains may 
be multidispersed having random size, shape and orientation. By using results from integral 
geometry, the MEC of patterns formed by clusters of overlapping grains can be  
calculated exactly \citep{Mecke_91}. In $d= 2$ the MEC reads 
\begin{equation}
\label{eq:mec_cont_2d}
\mecb_2(\eta)=\eta(I-\eta)e^{-\eta}, \quad \eta=\rho a,
\end{equation}
with the grain density $\rho$, and the dimensionless ratio $ I= \frac{u}{4\pi a}$; $a$ and $u$ 
denote the average grain area and perimeter, both assumed to be distributed 
independently from the grain locations. In the case of monodispersed convex grains, $I \geq 1$ is 
an isoperimetric ratio. For $d = 3$ one finds 
\begin{equation}
\mecb_3(\zeta)=(1-3I_1 \zeta+ \frac{3\pi^2}{32}I_2 \zeta^2)e^{-\zeta}, \quad \zeta=\rho v.
\end{equation}
Here, $I_1=\frac{sb}{6v}$, $I_2=\frac{s^3}{36\pi v^2}$; $b$, $s$ and 
$v$ are the averages of the grain mean breadth, surface area and volume, respectively.
Again, for monodispersed convex grains, $I_1\geq 1$, $I_2 \geq 1$. 
The graphs of $\mecb(\pb)$ and $\mecw(\pb)$, when plotted as functions of the mean coverage  
$0\leq 1-e^{-\eta} < 1$ and $0\leq1-e^{-\zeta} <1$ are similar to the graphs of the 
corresponding lattice MECs: $\mecb_2(\eta)$ has a single zero at $\eta_0=I^{-1}$, 
$\mecb_3(\zeta)$ has two zeros located at 
$\zeta_0=\frac{96}{\pi}\frac{bv}{s^2}\left(1\pm \sqrt{1-\frac{\pi}{24}\frac{s}{b^2}}\right)$.
For monodispersed discs $\eta_0= 1$, which may be compared 
with the threshold value $\eta_c\approx1.12$. Monodispersed spheres yield $\zeta_0^-=0.377$, 
to be compared with the percolation threshold of penetrable spheres, $\zeta_c\approx0.34$ \citep{Lorenz_JChemPhys_01}. The 
interpretation of the continuum MECs in terms of cluster structures can be carried over 
unchanged from the lattice examples. Thus, for instance, $\zeta_0^+$ is expected to provide 
a quick estimate for the percolation threshold $\zeta_c^+$ of the void space defined 
as the set complement of the pattern by the clusters of overlapping grains.  

\section{Summary and Discussion}
Taken together, the equations (\ref{eq:ec_crit_percolation}) and (\ref{eq:pc_estimation}) represent our main result. The relation (\ref{eq:pc_estimation}) 
connects the universal scaling behaviour of  percolating clusters at $\pc$ with their 
specific topological structure, as expressed by the mean Euler characteristic. It 
provides a novel and parameter-free estimate of threshold values for two-dimensional lattices, which is
remarkably accurate. In the case of self-matching lattices Eq.(\ref{eq:pc_estimation}) 
reproduces the exact values $\pc = 1/2$, and offers an explanation for the numerical 
finding that the zero crossing of the MEC, $\pz$, is a tight upper (lower) bound 
for  the threshold $\pc$ if $\pc > 1/2$ ($\pc < 1/2$). 

Over the years, a variety of ``universal'' approximate formulae for predicting 
percolation thresholds have been devised \citep{Scher_70, Suding_PRE_99, Galam_PRE_96, Wierman_PRE_05b}. 
Most of these proposals fit an empirical relation with a number of free parameters
to a set of known thresholds. 
Recently, \citeauthor{Wierman_PRE_05a} \citep{Wierman_PRE_05a} introduced a list of criteria for the 
evaluation of such formulae. Accordingly, these should (1) be well defined,  
(2) be easily computable, (3) provide values only between 0 and 1, (4) depend 
only on the adjacency structure of the lattice, (5) be accurate, (6) be consistent 
with the matching relationship, and (7) be consistent with the containment principle. 

The Eq.(\ref{eq:pc_estimation}) complies with the first six of these requirements, as can be inferred 
from its deduction and from the comparision of our estimates with precise 
numerical threshold values. We have not checked in detail the criterion (7). 
The claim, that our $\pc$-estimation is well defined may perhaps be questioned, 
since Eq.(\ref{eq:pc_estimation}) involves an extrapolation of the scaling ansatz to the smallest 
cluster sizes. However, this ad hoc simplification can be removed by going back 
to the ``master equation'' (\ref{eq:ec_crit_percolation}), which provides means to systematically correct 
for small-scale irregularities, as shown in \TAB{small_clusters}. 

We mentioned the exact expressions of the mean Euler characteristic for 
3d-cubic lattices and for the Boolean model of continuum percolation in two and 
three dimensions. The close linking of the zero crossings of these MECs to the respective threshold densities appears 
to persist and underlines once more the role of the Euler characteristic as the 
appropriate concept for describing the topological aspects of percolation. But to 
apply this intriguing fact for the construction of threshold estimators for three 
dimensions and for continuum models remains a challenging problem.

\ack
We would like to thank Robert Ziff for valuable comments on the manuscript.
RAN acknowledges financial support by the DFG and the state of Bavaria via IDK-NBT.

\end{document}